\documentclass[doublespacing]{elsart}

\usepackage{natbib}
\usepackage{graphicx}

\usepackage{amssymb}

\begin{document}
\begin{frontmatter}



\title{Efficiency and Current in a correlated ratchet}


\author[a]{Bao-Quan  Ai},\author[a]{Guo-Tao  Liu},
\author[b]{Hui-Zhang Xie}, \author[a,b]{De-Hua Wen}, \author[c]{Xian-Ju Wang}, \author[d]{Wei Chen}
 and \author[a]{Liang-Gang Liu}
\address[a]{Department of Physics, ZhongShan University,
GuangZhou, China}
\address[b]{Department of Physics, South China University of
technology, GuangZhou, China}
\address[c]{Department of Physics, South China Normal University, GuangZhou, China}
\address[d]{Department of Physics, JiNan  University, GuangZhou,
China}
\begin{abstract}
\baselineskip 0.2in \indent We present a detailed study of the
transport and the efficiency of a ratchet system in a periodic
potential in the presence of correlated noises.
 The current and the efficiency of the system are investigated. It
is found that, when the potential is spatially symmetric, the
correlation between the two noises can induce a net transport. The
efficiency shows many interesting features as a function of the
applied force, the noise intensity, the external load, etc. The
efficiency can be maximized as a function of noise intensity (or
temperature), which shows that the thermal fluctuation can
facilitate the efficiency of energy transformation.
\end{abstract}

\end{frontmatter}

\textbf{In the presence of correlated noises, the current and the
efficiency of a ratchet system in a periodic potential are
investigated. It is found that the correlation between the two
noises can induce a net transport when the potential is spatially
symmetric. The efficiency can be maximized as a function of the
noise intensity, which shows that the thermal fluctuation can
facilitate the efficiency of the energy transformation.}

\baselineskip 0.3in
\section {Introduction}

\indent Much of the interest in non-equilibrium-induced transport
processes is concentrated on stochastically driven ratchets
\cite{1}\cite{2}\cite{3}. This subject was motivated by the
challenge to explain the origin of directional transport in
biological systems. The rectification of noise leading to
unidirectional motion in ratchet systems has been an active field
of research over the last decade. In these systems the directed
Brownian motion of particles is induced by nonequilibrium noise in
the absence of any net macroscopic forces and potential gradients.
Several physical models have been proposed: rocking ratchets
\cite{4}\cite{5}\cite{6},  flashing ratchets \cite{7}, diffusion
ratchets \cite{8}, correlation ratchets \cite{9} and deterministic
ratchets \cite{10}\cite{11}. In all these studies the potential is
taken to be asymmetric in space. It has also been shown that one
can obtain a unidirectional current in the presence of spatially
asymmetric potentials. For these nonequilibrium systems an
external force should be asymmetric or the presence of space
dependent mobility is required.

 \indent
Molecular motors are known to have the high efficiency of
 energy transformation in the presence of thermal fluctuation
 \cite{1}. Recently, motivated by the surprising fact,  the
 efficiency with which the ratchet converts fluctuation to useful
 work is a much interest subject. New questions
 regarding the nature of heat engines at molecular scales are
 being investigated \cite{12}.

\indent In most of previous investigations, various noises are
assumed to have different origins and are treated as independent
random variables \cite{13}. However, in certain situations these
noises may have a common origin and thus may be correlated with
each other \cite{14}\cite{15}\cite{16}.  The correlations between
the noises can change the properties of stochastic systems. In
this paper, a novel condition of the correlated ratchet operation
is investigated where the potential is spatially symmetric, namely
neither spatially asymmetry in the potential  nor temporal
asymmetry in the driving force is required, the correlation
between the two noises can induce transport. The efficiency of the
ratchet is also presented in a periodic potential in the presence
of an adiabatic external period force. The organization of the
paper is as follows: In Sec. 2 we introduce the model and the
basic quantities, namely, the average probability current and the
efficiency of the ratchet. In Sec. 3 we discuss the current of the
ratchet with the symmetric potential for the case of no external
load. Sec. 4 is devoted to the exploration of the efficiency of
the system. The summary
and discussion of our findings is presented in Sec. 5.\\

\section {The Forced Thermal Ratchet with Correlated Noises}
\indent We consider a forced ratchet system subject to an external
load and the equation of motion of the ratchet reads
\begin{equation}\label{1}
m\frac{d^{2}x}{d t^{2}}=-\beta \frac{d x}{d t}-\frac{d V_{0}(x)}{d
x}+a_{0}F(t)+\xi_{2}(t)F(t)+\xi_{1}(t)-\frac{d V_{L}(x)}{d x},
\end{equation}
where $x$ represents the position of Brownian particle, $m$
denotes the mass of the particle, $\beta$ is the viscous friction
drag coefficient, $F(t)$ is an external periodic force,
$F(t+\tau)=F(t)$, $\int^{\tau}_{0}F(t)dt=0$ and $F_{0}$ is
amplitude of $F(t)$, see Fig.1 .
$V_{0}(x+2n\pi)=V_{0}(x)=-\sin(x)$, $n$ is any natural integer and
$V_{L}(x)$ is a potential against which the work is done and
$\frac{d V_{L}(x)}{d x}=L>0$. $\xi_{1}(t)$, $\xi_{2}(t)$ are white
noises with zero mean.

\indent Because the motion of the ratchet is highly overdamped,
 the inertia term can be neglected. Hence, thereafter in the place
 of Eq. (1) we shall make use of the following equation in the
 case of $\beta=1$ and $a_{0}=1$
\begin{equation}\label{1}
\frac{d x}{d t}=-\frac{d V_{0}(x)}{d
x}+F(t)+\xi_{2}(t)F(t)+\xi_{1}(t)-\frac{d V_{L}(x)}{d x}.
\end{equation}

\indent In some situations various noise sources must be
considered. For example, Millonas \cite{13}investigated a ratchet
system including a subsystem, a thermal bath and a nonequilibrium
bath that contains two parts: one is the nonthermal part which can
be view as a source, the other is thermal part. Here it is easy to
see that the two noises are treated as independent random
variables. However, in some situations these noises may have a
common origin and thus may be correlated with each other. In the
present work, we assume that the two noises $\xi_{1}(t)$,
$\xi_{2}(t)$ are correlated with each other and the correlations
between the noises have the following form \cite{16}\cite{17}
\begin{equation}\label{2}
  <\xi_{i}(t)\xi_{j}(t^{'})>=2C_{i,j}\sqrt{D_{i}D_{j}}\delta(t-t^{'}),
\end{equation}
where $C_{i,j}=\lambda$ for $i\neq j$ and $ C_{i,j}=1$ for $i=j$,
$\lambda$ denotes the cross-correlation degree between
$\xi_{1}(t)$ and $\xi_{2}(t)$, and $-1 \leq \lambda \leq 1$. \\
\indent When $F(t)$ changes slowly enough, the system could be
treated as quasistatic, the evolution of probability density
$P(x,t)$ described by Eq. (2) is then given by the corresponding
Fokker-Planck equation,
\begin{equation}\label{3}
\frac{\partial P(x,t)}{\partial t}=\frac{\partial}{\partial
x}[U^{'}(x,F_{0})+G(F_{0},\lambda)\frac{\partial}{\partial
x}]P(x,t)=-\frac{\partial j(x,t)}{\partial x},
\end{equation}
where the probability current density $j(x,t)$ is given by
\begin{equation}\label{4}
j(x,t)=-U^{'}(x)P(x,t)-G(F_{0},\lambda)\frac{\partial
P(x,t)}{\partial x},
\end{equation}
where $U^{'}$ denotes  the first derivative of $U(x)$, and
\begin{equation}\label{5}
U(x)=V_{0}(x)+V_{L}(x)-F_{0}x=-\sin(x)+Lx-F_{0}x,
\end{equation}
\begin{equation}\label{6}
G(F_{0},\lambda)=D_{2}F_{0}^{2}+2\lambda
F_{0}\sqrt{D_{1}D_{2}}+D_{1}.
\end{equation}
\indent In a quasi stationary state, the steady current of the
particle can be solved by evaluating the constants of integration
under the normalization condition and the periodicity
 condition of $P(x)$, the current can be obtained and expressed as \cite{16}
\begin{equation}\label{7}
  j=\frac{G(F_{0},\lambda)[1-\exp(-2\pi(F_{0}-L)/G(F_{0},\lambda))]}{\int^{2\pi}_{0}{\exp[\phi(x)]dx } \int^{x+2\pi}_{x}{\exp[-\phi(y)]dy
  }},
\end{equation}
where the generalized potential is given by
\begin{equation}\label{8a}
\phi(x)=-\frac{U(x)}{G(F_{0},\lambda)}=\frac{\sin(x)+(F_{0}-L)x}{D_{2}F_{0}^{2}+2\lambda
F_{0}\sqrt{D_{1}D_{2}}+D_{1}}.
\end{equation}
The average probability current over the time interval of a period
can be expressed,
\begin{equation}\label{8}
  <j>=\frac{1}{2}[j(F_{0})+j(-F_{0})].
\end{equation}
\indent The input energy $R$ per unit time from an external force
to the ratchet and the work $W$ per unit time that the ratchet
system extracts from the fluctuation into the work are given,
respectively \cite{18},
\begin{equation}\label{9}
R=\frac{1}{t_{j}-t_{i}}\int^{x(t_{j})}_{x(t_{i})}F(t)dx(t),
\end{equation}
\begin{equation}\label{10}
W=\frac{1}{t_{j}-t_{i}}\int^{x(t_{j})}_{x(t_{i})}dV(x(t)),
\end{equation}
where $V(x(t))=V_{0}(x(t))+V_{L}(x(t))$.

\begin{figure}[htbp]
  \begin{center}\includegraphics[width=10cm,height=7cm]{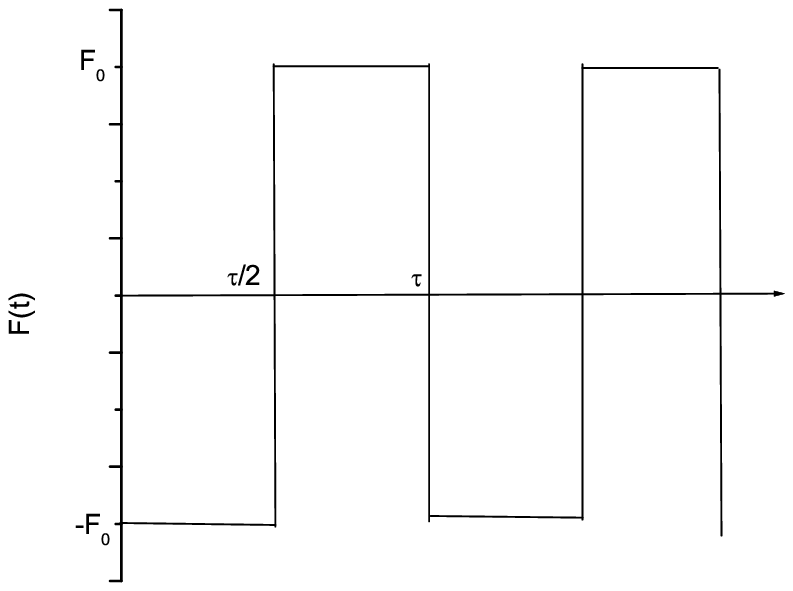}
  \caption{The driving force $F(t)$, $<F(t)>=0$, $F(t+\tau)=F(t)$, and $F_{0}$ is amplitude.}\label{1}
\end{center}
\end{figure}
 For a quasi-static force,  they yield
\begin{equation}\label{11}
  <R>=\frac{1}{2}F_{0}[(j(F_{0})-j(-F_{0}))],
\end{equation}
\begin{equation}\label{12}
<W>=\frac{1}{2}L[j(F_{0})+j(-F_{0})].
\end{equation}
\indent Thus the efficiency $\eta$ of the system to transform the
external energy to useful work is given by
\begin{equation}\label{13}
  \eta=\frac{<W>}{<R>}=\frac{L[j(F_{0})+j(-F_{0})]}{F_{0}[j(F_{0})-j(-F_{0})]}.
\end{equation}
\section{Current in the system for the case $L=0$ }
\indent From Eq. (8) it may be noted that even for $L=0$,
 $j(F_{0})$ may not be equal to $-j(-F_{0})$ for $\lambda \neq 0$,
so the average current $<j>\neq 0$. The fact leads to the
rectification of current in the presence of an applied force
$F(t)$. Based on Eq. (10), the results are given by Fig. 2-Fig. 5.\\

\begin{figure}[htbp]
  \begin{center}\includegraphics[width=10cm,height=7cm]{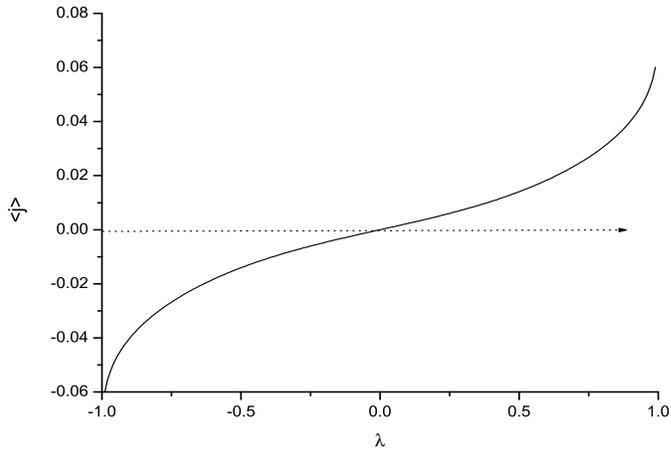}
  \caption{The current versus $\lambda$ for
symmetric potential. $D_{1}=0.5$,$D_{2}=0.5$, $F_{0}=1$, and
$L=0$.}\label{1}
\end{center}
\end{figure}

\indent In Fig. 2 we plot the curve of the current which is a
function of the intensity $\lambda$ of correlations between the
two noises. It is found that the critical value $\lambda_{c}=0$
for current $<j>=0$, the current is positive for $\lambda >0$ and
negative for $\lambda<0$. The current increases with $\lambda$. We
can see that the current reversal can occur at the case of  no
correlation and the ratchet can exhibit a current in either
direction. The correlation intensity in a symmetric potential case
play a
important role in the fluctuation-induced transport.\\

\begin{figure}[htbp]
  \begin{center}\includegraphics[width=10cm,height=7cm]{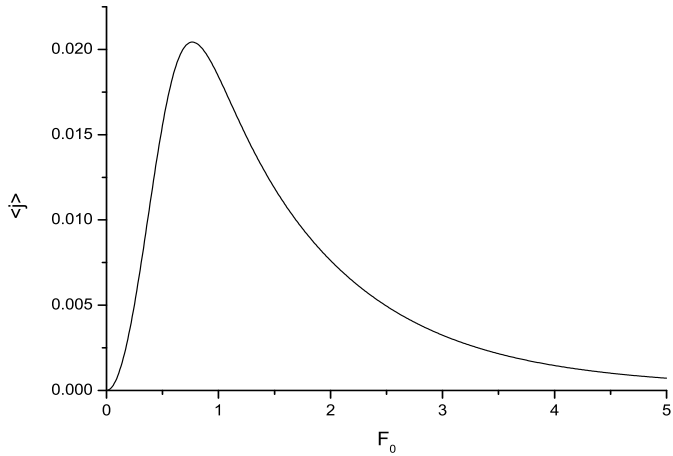}
  \caption{The current versus $F_{0}$ for symmetric potential.
$D_{1}=0.5$, $D_{2}=0.5$, $\lambda=0.5$ and $L=0$.}\label{2}
\end{center}
\end{figure}
\indent In Fig. 3, we plot the curve of the current versus the
amplitude of the adiabatic forcing $F_{0}$. It can be seen from
the figure that the current $<j>$ gives a maximum and saturates to
the zero value
in the large amplitude limit.\\
\begin{figure}[htbp]
  \begin{center}\includegraphics[width=10cm,height=7cm]{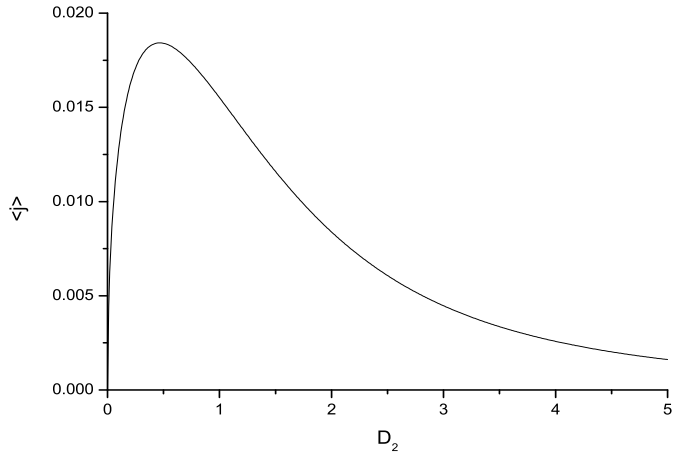}
  \caption{The current versus $D_{2}$ for symmetric potential.
$D_{1}=0.5$, $F_{0}=1$, $\lambda=0.5$ and $L=0$.}\label{3}
\end{center}
\end{figure}
\begin{figure}[htbp]
  \begin{center}\includegraphics[width=10cm,height=7cm]{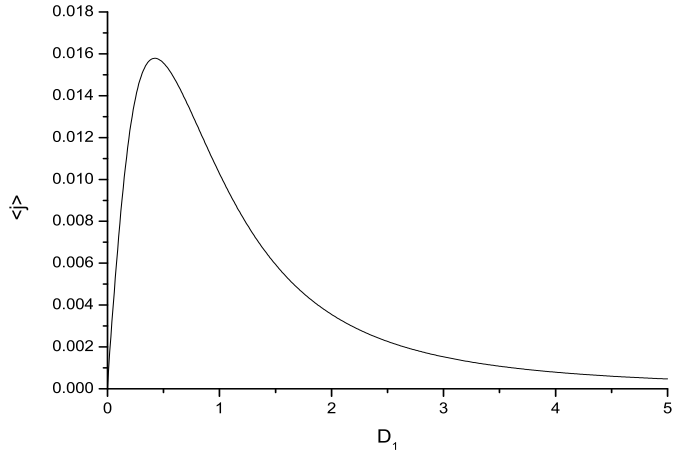}
  \caption{The current versus $D_{1}$ for symmetric potential.
$D_{2}=0.5$, $F_{0}=0.5$, $\lambda=0.5$, and $L=0$.}\label{4}
\end{center}
\end{figure}

\indent Fig. 4 shows that the current is a peaked function of
$D_{2}$ and the current goes to zero for a large noise case.
Similarly, the current is a peaked function of $D_{1}$ as shown in
Fig. 5. The results shows that an optimal transport occurs at some
noise cases and there is no net current for a large noise case.\\

\indent In our ratchet the potential of the ratchet is completely
symmetric in space and the external driving force $F(t)$ is
completely symmetric in time. In a period of external force
$F(t)$, the effective generalized potential
$\phi_{eff}(x)=1/\tau\int_{0}^{\tau}\phi(x,F(t))dt=[\phi(x,F_{0})+\phi(x,-F_{0})]/2$,
so
\begin{equation}\label{2Q}
\phi_{eff}(x,\lambda)=\frac{1}{2}[\frac{\sin(x)+F_{0}x}{D_{2}F_{0}^{2}+2\lambda
F_{0}\sqrt{D_{1}D_{2}}+D_{1}}+\frac{\sin(x)-F_{0}x}{D_{2}F_{0}^{2}-2\lambda
F_{0}\sqrt{D_{1}D_{2}}+D_{1}}].
\end{equation}
Let us introduce $\Delta\phi_{eff}(\lambda)$to measure the slope
of $\phi_{eff}(x,\lambda)$ over a period
\begin{eqnarray}\label{1}
\Delta\phi_{eff}(\lambda)&=&\frac{\phi_{eff}(x+2\pi,\lambda)-\phi_{eff}(x,\lambda)}{2\pi} \nonumber\\
&=&\frac{-4\lambda F_{0}^{2}
\sqrt{D_{1}D_{2}}}{(D_{2}F_{0}^{2}+D_{1})^{2}-4\lambda^{2}F_{0}^{2}D_{1}D_{2}}.
\end{eqnarray}
It is easy to obtain
$\Delta\phi_{eff}(-\lambda)=-\Delta\phi_{eff}(\lambda)$, which
shows that $<j(-\lambda)>=-<j(\lambda)>$(see Fig.2 ). When there
is no correlations ($\lambda=0$)between noises, the effective
spatially generalized potential
$\phi_{eff}(x)=\frac{\sin(x)}{D_{2}F_{0}^{2}+D_{1}}$ is symmetric
and $\Delta\phi_{eff}(\lambda)=0$ (see Eq. (16) and Eq. (17)), so
no currents occur. However, From the Eq. (16), if $\lambda\neq 0$,
the effective potential is not the same as the symmetric potential
$V(x)$, since the two effects of external driving force in its
period can not be cancelled because of $\lambda\neq 0$, so the
symmetry of the effective generalized potential is broken by the
correlation between noises. It is known that the state-dependent
diffusion can induce transport in case of the symmetry of
generalized potential being broken. Therefore, correlations
between the two noises can induce transport. Some previous works
also involved a symmetry breaking. Reimann \cite{19} introduced a
model of interacting Brownian particles in a symmetric, periodic
potential that undergoes non-equilibrium phase transition. The
associated spontaneous symmetry breaking entails a ratchet-like
transport mechanism. The ratchet-like transport mechanism arise
through a symmetry breaking was also investigated by Mangioni
\cite{20} under a system of periodically coupled nonlinear phase
oscillators submitted to multiplicative white noises. In Buceta's
study \cite{21} the combination with local potential and external
fluctuations causes a symmetry breaking.

\section{Efficiency in Correlated Ratchet}
\indent In this section we discuss the efficiency and the
corresponding current of the ratchet in the presence of an
external load based on Eq. (13)-Eq. (15). Because
$\frac{j(-F_{0})}{j(F_{0})}<0$, Eq. (15) can be rewritten as
follow
\begin{equation}\label{14}
\eta=\frac{L}{F_{0}}\{1-\frac{2|\frac{j(-F_{0})}{j(F_{0})}|}{1+|\frac{j(-F_{0})}{j(F_{0})}|}\}.
\end{equation}
In the limit $|\frac{j(-F_{0})}{j(F_{0})}|\rightarrow 0$, the
maximum efficiency of the energy transformation for given load $L$
and force amplitude $F_{0}$ is given:
$\eta_{max}=\frac{L}{F_{0}}$. The results are represented by Fig.
6-Fig. 9.\\

\begin{figure}[htbp]
  \begin{center}\includegraphics[width=10cm,height=7cm]{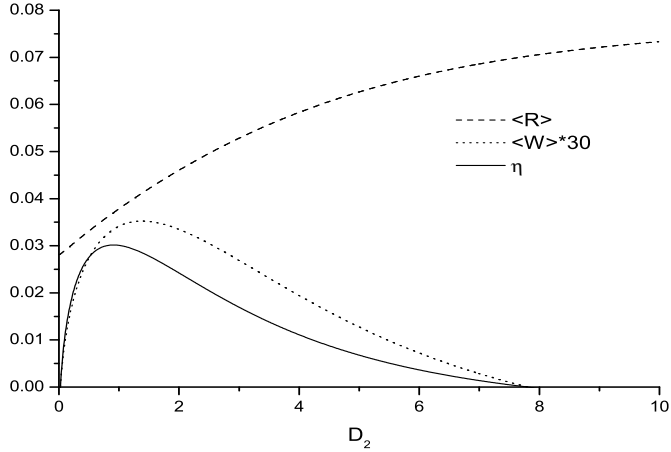}
  \caption{Efficiency $\eta$, $<R>$, $<W>$ as a function of $D_{2}$
for $\lambda=0.5$. $F_{0}=0.5$, $D_{1}=0.5$, $L=0.04$. $<W>$ has
been scaled up by a factor 30 to make it comparable with $\eta$
and $<R>$. Y-axis is in dimensionless units.}\label{5}
\end{center}
\end{figure}
\indent In Fig. 6 we plot the efficiency $\eta$, input energy
$<R>$ and work done $<W>$ (scaled up by a factor 30 for
convenience of comparison) as a function of $D_{2}$ for the
parameter values, $F_{0}=0$, $D_{1}=0.5$, $\lambda=0.5$, $L=0.04$.
The figure shows that the efficiency exhibits a maximum as a
function of $D_{2}$ which indicates that the thermal fluctuation
facilitates energy conversion. The input energy $<R>$ increases
with $D_{2}$ monotonically and saturates to a certain value in a
large noise intensity limit. The work $<W>$ exhibits a
maximum as a function of $D_{2}$.\\

\begin{figure}[htbp]
  \begin{center}\includegraphics[width=10cm,height=7cm]{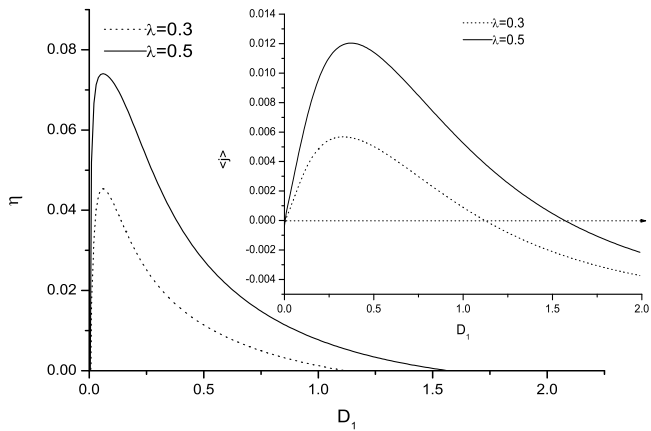}
  \caption{Efficiency $\eta$ as function of $D_{1}$ for different
values of correlation $\lambda$=0.3, 0.5. $D_{2}=0.5$,
$F_{0}=0.5$, $L=0.04$. The inset shows the $<j>$ as a function of
$D_{1}$ for the same parameters.}\label{6}
\end{center}
\end{figure}
\indent Fig. 7 shows the efficiency $\eta$ as a function of
$D_{1}$ for different value of $\lambda$. The efficiency is a
peaked function of $D_{1}$ and increases with the parameter
$\lambda$. Similarly, the current is a peaked function of $D_{1}$
for the corresponding parameters as shown in the inset. From the
inset we can notice that the current reverses its direction for
the case $\lambda\neq 0$. It is noted that noise intensity $D_{1}$
corresponding to maximum efficiency is not the same as the noise
intensity at which the current $<j>$ (see the inset of Fig. 7) is
maximum. The difference is attributed to the observation that the
efficiency is ratio of the extracted $<W>$ to the consumed energy
$<R>$. The work $<W>$ is purely proportional to the current $<j>$.
However, the consumed energy is not a constant but varies
sensitively according to the condition. Therefore the efficiency
$\eta$ is not simply proportional to the current $<j>$.

\indent In Fig. 6 and Fig. 7 we obtained the resonance curve which
is the same as the previous results \cite {22}. When $D_{1},
D_{2}\rightarrow 0$, thermal diffusion over the potential barriers
vanishes, no net currents occur, so $\eta\rightarrow 0$. On the
other hand, when $D_{1}, D_{2}\rightarrow \infty$, diffusion
becomes insensitive to the ratchet amplitude and its modulation,
no net currents occur, too, so $\eta\rightarrow 0$. The resonance
occurs at a finite noise intensity.

\begin{figure}[htbp]
  \begin{center}\includegraphics[width=10cm,height=7cm]{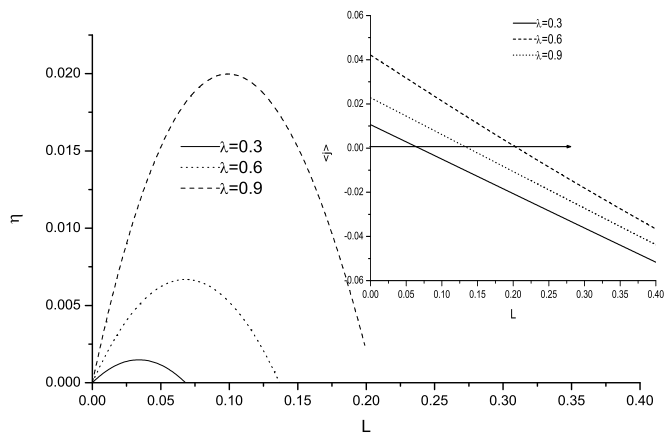}
  \caption{Efficiency $\eta$ as function of $L$ for different values
of correlation $\lambda$=0.3, 0.6, 0.9. $D_{1}=0.5$, $D_{2}=0.5$,
$F_{0}=0.5$. The inset shows the $<j>$ as a function of $L$ for
the same parameters.}\label{7}
\end{center}
\end{figure}
\indent In Fig. 8, we plot the efficiency $\eta$ (the inset for
current $<j>$) versus load $L$ for different values of $\lambda$.
It is expected that the efficiency also exhibits a maximum as a
function of the load. It is obvious that the efficiency is zero
when load is zero. At the critical value $L_{c}$ (beyond which
current flows in the direction of the load (see the inset of
Fig.8)) the current is zero and hence the efficiency vanishes
again. Between the two extreme values of the load the efficiency
exhibits a maximum. Beyond $L=L_{c}$ the current flows down the
load, and therefore,
the definition of efficiency becomes invalid.\\

\begin{figure}[htbp]
  \begin{center}\includegraphics[width=10cm,height=7cm]{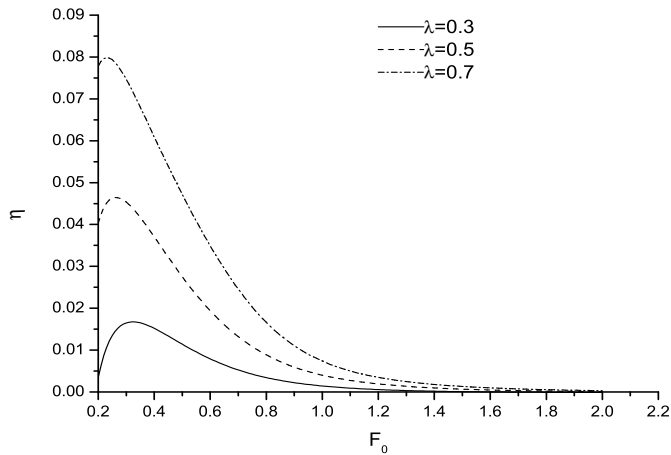}
  \caption{Efficiency $\eta$ as function of $F_{0}$ for different
values of correlation $\lambda$=0.3, 0.5, 0.7. $D_{1}=0.5$,
$D_{2}=0.5$, $L=0.04$.}\label{8}
\end{center}
\end{figure}
\indent The efficiency $\eta$ versus the amplitude of the
adiabatic forcing $F_{0}$ for different values of $\lambda$ is
shown in Fig. 9. It can be seen from the figure that efficiency
exhibits a maximum and saturates to the same value (zero) in large
amplitude limit. Beyond the large amplitude limit the efficiency
increases with $\lambda$,  which indicates that the correlation
between the two noises plays a important role for
energy transformation.\\

\section {Summary and Conclusion}
\indent In the present paper, we study the current and the
efficiency of a forced thermal ratchet with correlated noises. For
the case of no external load, no net current occurs when there is
no correlation ($\lambda=0$) between the thermal noises, while the
net current occurs for the case of $\lambda \neq 0$. It is obvious
that the symmetry of effective generalized potential is broken by
the correlation between the noises. Therefore, neither spatially
asymmetry nor temporal asymmetry is required, the correlation can
induce a net transport. The current is a peaked function of
$D_{2}$, $D_{1}$ and $F_{0}$ and goes to zero for the case of
large
$D_{2}$, $D_{1}$ and $F_{0}$ limit.\\

\indent For the case of $L\neq 0$, the efficiency of the ratchet
is investigated. The efficiency shows a maximum as a function of
noise intensity as does the net current, though they do not occur
at the same noise intensity. It is obvious that the thermal
fluctuation is not harmful for fluctuation-induced work even
facilitates its efficiency. The current reversal occurs at
$\lambda\neq 0$ which is different from
the case of $L=0$.\\

\indent Based on energetic analysis of the ratchet model Kamegawa
et al. \cite{18} made an important conclusion that the efficiency
of energy transformation cannot be optimized at finite
temperature. However, the discussion in that paper was only on the
quasistatic limit where the change of the external of force is
slow enough. Takagi and Hondou\cite{23} found that thermal noise
may facilitate the energy conversion in the forced thermal ratchet
when the ratchet is not quasistatic.  Recently, investigation of
Dan et al. \cite{24} showed that the efficiency can be optimized
at finite temperatures in inhomogeneous systems with spatially
varying friction coefficient in an adiabatically rocked ratchet.
Efficiency optimization in homogeneous non-adiabatic ratchet
systems was observed by Sumithra et al.\cite{25}. Sokolov and
Blumen \cite{26} found that the maximal efficiency is obtained at
a finite temperature in a ratchet which consists of an array of
three-level systems placed sequentially. In our system the thermal
fluctuations actually facilitate the energy transformation under
the homogeneous adiabatic condition which is not the same as the
previous works.

{\bf Acknowledgements}\\
 \indent The work was supported by National
Natural Science Foundation of China (Grant No. of 10275099) and
GuangDong Provincial Natural Science Foundation (Grant No. of
021707 and 001182).\\



\end{document}